\documentclass[prl,twocolumn,,aps,showpacs]{revtex4} % showkeys

\usepackage{graphicx}% Include figure files
\usepackage{dcolumn}% Align table columns on decimal point
\usepackage{bm}% bold math
\usepackage{amssymb}
\usepackage{url}
\PassOptionsToPackage{hyphens}{url}\usepackage{hyperref}
\usepackage{color}
\begin{document}

\title{Forensic analysis of the Turkey 2023 presidential election reveals extreme vote swings in remote areas}

\author{Peter Klimek$^{1,2,3*}$, Ahmet Ayka\c{c}$^{4}$, Stefan Thurner$^{1,2,3,5}$}
\affiliation{$^1$Section for Science of Complex Systems, CeDAS, Medical University of Vienna, Spitalgasse 23, A-1090, Austria\\
$^2$Complexity Science Hub Vienna, Josefst\"adter Strasse 39, A-1080 Vienna, Austria\\
$^3$Supply Chain Intelligence Institute Austria (ASCII), Josefst\"adter Strasse 39, A-1080 Vienna, Austria\\
$^4$Theseus International Management Institute, Sophia Antipolis, France\\
$^5$Santa Fe Institute, 1399 Hyde Park Road, Santa Fe, NM 85701, USA\\
$^*$\href{mailto:peter.klimek@meduniwien.ac.at}{peter.klimek@meduniwien.ac.at} }
\date{\today}

\begin{abstract}
Concerns about the integrity of Turkey's elections have increased with the recent transition from a parliamentary democracy to an executive presidency under Recep Tayyip Erdo\u{g}an. 
Election forensics tools are used to identify statistical traces of certain types of electoral fraud, providing important information about the integrity and validity of democratic elections. 
Such analyses of the 2017 and 2018 Turkish elections revealed that malpractices such as ballot stuffing or voter manipulation may indeed have played a significant role in determining the election results. Here, we apply election forensic statistical tests for ballot stuffing and voter manipulation to the results of the 2023 presidential election in Turkey. 
We find that both rounds of the 2023 presidential election exhibit similar statistical irregularities to those observed in the 2018 presidential election, however the magnitude of these distortions has decreased. 
We estimate that 2.4\% (SD 1.9\%) and 1.9\% (SD 1.7\%) of electoral units may have been affected by ballot stuffing practices in favour of Erdo\u{g}an in the first and second rounds, respectively, compared to 8.5\% (SD 3.9\%) in 2018. 
Areas with smaller polling stations and fewer ballot boxes had significantly inflated votes and turnout, again, in favor of Erdo\u{g}an. 
Furthermore, electoral districts with two or fewer ballot boxes were more likely to show large swings in vote shares in favour of Erdo\u{g}an from the first to the second round.
Based on a statistical model, we estimate that these swings translate into 342,000 excess votes (SD 4,900) or 0.64\% for Erdo\u{g}an.   
Our results suggest that Turkish elections continue to be riddled with statistical irregularities, that may be indicative of electoral fraud.
\end{abstract}

%\pacs{}% PACS

\maketitle

The first round of the 2023 presidential election was held on 14 May 2023 and pitted incumbent President Recep Tayyip Erdo\u{g}an of the Justice and Development Party (AKP) against opposition candidate Kemal Kili\c{c}daro\u{g}lu, who led an alliance of six opposition parties. The election was seen as the first real threat to Erdo\u{g}an's presidency in a long time, as Turkey was reeling from a prolonged economic crisis with inflation rates of up to 85\% and the aftermath of devastating earthquakes in February 2023 that killed more than 50,000 people, coupled with public outrage at the government's slow response to these crises \cite{WaPo23}. Despite polls to the contrary, Erdo\u{g}an won by a healthy margin over Kili\c{c}daro\u{g}lu with 49 per cent to 45 per cent of the vote, sending Turkey into a run-off between the two candidates on 28 May. Erdo\u{g}an won this second round of the election with a margin of 4.36\% (52.18\% versus 47.82\%) over Kili\c{c}daro\u{g}lu.

While elections in Turkey are generally considered to be free and fair, the playing field is clearly not level. For example, according to Turkish media monitoring, Erdo\u{g}an received almost 33 hours of airtime on the main state television channel, compared to 32 minutes for Kili\c{c}daro\u{g}lu \cite{APN}. These and other imbalances became possible after Erdo\u{g}an transformed Turkey from a parliamentary democracy to an executive presidency following a 2017 referendum in which a reform package narrowly won. That election, however, was marred by allegations of fraud, as unverified videos and reports emerged on social media showing various forms of electoral malpractice, such as ballot stuffing (casting multiple votes for one candidate) and voter coercion (preventing potentially opposing voters from casting their ballots) \cite{Observer17, NYT}. Also practices such as handing out cash to supporters have been reported \cite{Guardian}.

The emerging field of electoral forensics seeks to diagnose the extent to which a particular type of malpractice may have affected the outcome of an election, in order to identify electoral malpractice in a timely and fully quantitative manner \cite{Levin12}.
A disproportionate abundance of round numbers was often the focus of early work in election forensics. 
The basic principle of these tests is that humans have a particular tendency to favour round numbers, or numbers with certain digits, when producing results. 
These tendencies are at odds with the statistics of the expected number and digit distributions of fair elections \cite{Cantu11,Beber12}, including violations of Benford's Law \cite{Mebane08, Pericchi11,Gueron22}.
However, it has been shown that such digit-based tests need to be combined with contextual information such as country-specific risk factors, socio-economic inequalities or ethnic fractionalisation \cite{Montgomery15}.

As a result, there has been a growing interest in forensic electoral testing that attempts to provide ``mechanistic'' or generative models of the impact of specific electoral malpractices on the expected distributions of votes and turnout across polling stations, as well as on the correlations between votes and turnout \cite{Myagkov09, Levin09, Klimek12, Jimenez17, Rozenas17, Jimenez11, Lacasa19}.
The basic rationale of such approaches is to consider elections as large-scale natural experiments in which a population is divided into a large number of electoral units in which each registered voter makes the decision to (i) cast a valid ballot or not, and (ii) vote for a particular candidate.
The large number of electoral units in most countries means that certain statistical regularities can be expected to hold.
Election forensics then tests whether deviations from these regularities are consistent with specific types of fraud. 
Similar principles can be used to apply machine learning models for election forensics \cite{Zhang19}.
These statistical tools are often complemented by analyses of secondary data, such as exit polls or survey and sampling data \cite{Prado11, Hausmann11}.

In a forensic analysis of the 2017 election, we confirmed that the election records indeed show specific statistical irregularities that point toward ballot stuffing and voter coercion \cite{Klimek18}. More specifically, ballot stuffing, when carried out on a large scale, typically leads to a correlation between voter turnout and the vote for the candidate \cite{Klimek12}. Voter manipulation or coercion tends to inflate votes and turnout in smaller or more remote regions, where opponents are easier to identify and irregularities are less likely to be observed or reported \cite{Jimenez17}.
For the constitutional referendum, we found that 11\% of areas were potentially affected by ballot stuffing and that removing the affected influences from the data would have turned the overall vote of the referendum from ``Yes'' to ``No''. There were also small but significant traces of ballot stuffing. Similar statistical irregularities were also observed in the 2018 presidential and parliamentary elections.

Here we ask whether similar forensic patterns are also present in both rounds of the 2023 presidential election \cite{YSK}. 
We test whether statistical fingerprints of ballot stuffing or voter coercion can also be found in the 2023 results.
We also seek to clarify the contribution of small and/or remote electoral units to such irregularities.
In particular, we develop a statistical test to compare the results at the polling station level for the first and second rounds of the presidential election.
With this test, we ask whether boxes in such areas are more likely to show improbably large swings of votes in favour of either candidate from one round to the next.

\section{Data and Methods}

We analyse the official and final election results of both rounds of the 2023 presidential election, as provided by the election commission \cite{YSK}.
At the finest level of aggregation available, the data for the first round consists of 191,872 electoral units (here also referred to as ``boxes'') for which we consider the size of the electorate (number of eligible voters), the number of valid votes and the number of votes for Recep T. Erdo\u{g}an.
The same variables were extracted from the data for the second round, which contained 192,214 electoral units.
Of these, 12 and 18 units respectively were removed because their number of valid votes was zero.

When applying the ballot-stuffing and voter-rigging tests, all electoral units with an electorate of less than one hundred voters are excluded in order to rule out that the results are driven by small number artefacts. This reduces the number of electoral units used to 180,301 and 180,629 respectively. 
The number of valid votes divided by the size of the electorate, $n_i^{(j)}$, for a ballot box $i$ is called the ``turnout'', $t_i^{(j)}$, for the first ($j=1$) or second ($j=2$) round of the election.
The number of votes, $V_i^{c,j}$, for candidate $c$ divided by the valid votes is called the ``vote share'', $v_i^{(c,j)}$ .

The election data includes multiple administrative levels for each ballot box, namely 81 provinces, 948 districts and 28,268 counties.
The number of ballot boxes is very unevenly distributed across these districts.
In particular, we identify districts that are typically smaller and more remote as those that have only two or fewer ballot boxes.

We apply ballot stuffing \cite{Klimek12} and voter coercion \cite{Jimenez17} to the data as previously reported \cite{Klimek18}.
In order to investigate whether there is systematic bias in vote shifts, we propose the following test procedure.
In the first round, Sinan O\u{g}an also ran and received 5.17\% of the vote, after which he supported Erdo\u{g}an in the run-off.
Therefore, we compare the vote shares of Erdo\u{g}an ($c=E$) or O\u{g}an ($c=O$) in round 1 with Erdo\u{g}an's shares in round 2. 
To do this, we compute the ``vote shift'' at the ballot box level as $\delta v_i = \frac{V_i^{(E,2)}}{t_i^{(2)}}-\frac{V_i^{(E,1)}+V_i^{(O,1)}}{t_i^{(1)}}$.
We assume that there is a general population-level trend in how preferences for a candidate may have shifted between the first and second rounds, which can be obtained as the mode of the distribution of $\delta v_i$, denoted as $\bar{\delta v}$.
Algorithmically, we estimate $\bar{\delta v}$ using MatLab's kernel density estimation procedure.
Our null hypothesis is that the vote shifts at the ballot box level are symmetrically distributed around this mode, i.e. the expectation $E$ for the deviations from $\bar{\delta v}$ is zero, $H_0: E(\delta v_i\bar{\delta v}) = 0$.
 
Let $B^{+/-}$ be the set of ballot boxes for which this deviation is greater/smaller than zero, $B^{+/-}=\{i | \delta v_i-\bar{\delta v} >/< 0 \}$.
A symmetrized vote shift distribution can then be constructed by replacing, for example, $(\delta v_i-\bar{\delta v})\forall i \in B^+$ by values of $-(\delta v_j-\bar{\delta v})$, where $j$ was randomly sampled from $B^-$. 
For the case were vote shifts typically favor one candidate, say $ E(|\delta v_i\bar{\delta v}|)_{B^+} > E(|\delta v_i\bar{\delta v}|)_{B^-}$ (meaning that the expectation value is taken over all $i$ in $B^{+/-}$), one replaces vales in $B^+$ with values sample from within $B^-$.  
This gives a model estimate for a symmetrized vote shift distribution from which corrected vote totals for individual candidates can be estimated.
By comparing the actual vote tallies with expectations from the model with symmetrized vot shift distributions, one obtains the number of excess votes due to large vote swings.

\section{Results}

\begin{figure}[tbp]
\begin{center}
 \includegraphics[width = 0.45\textwidth, keepaspectratio = true]{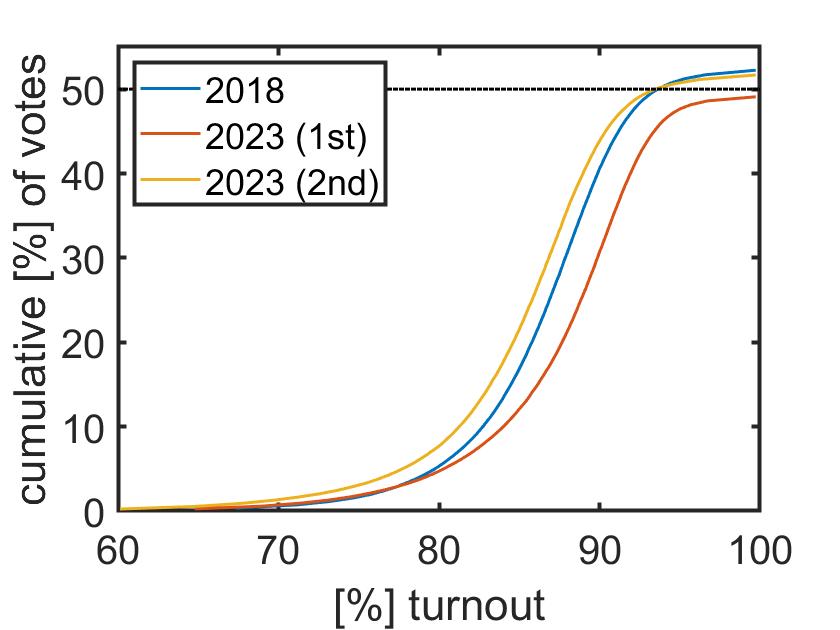}
\end{center}
 \caption{Votes for Erdo\u{g}an as a function of voter turnout for 2018 and 2023. For a given turnout level, the cumulative vote share of ballot boxes with this or lower turnout is shown. In 2018, a majority of more than 50\% is achieved by including boxes with a turnout of more than 90\% (blue). In the first round in 2023 (red line) we observe similar characteristics with higher turnout and lower vote shares (below 50\%), whereas in the second round (yellow) Erdo\u{g}an reached a majority of votes.}
 \label{CumVotes}
\end{figure}

The cumulative percentage of votes for Erdo\u{g}an is shown as a function of turnout in Figure~\ref{CumVotes}. For each turnout level (x-axis), the share of votes from boxes with that turnout level or lower is shown on the y-axis. In 2018, the share of votes exceeds the 50\% threshold only if we include voting boxes with a turnout of more than 90\%. In 2023, we observe a similarly shaped curve with an overall higher turnout (the curve is shifted to the right) but without crossing the 50\% threshold in the first round of the election. For the second round we find again a qualitatively similar curve as in 2018 that crosses the 50\% threshold at a turnout of around 92\%.

\subsection{Ballot stuffing test.}

\begin{figure*}[tbp]
\begin{center}
 \includegraphics[width = 0.9\textwidth, keepaspectratio = true]{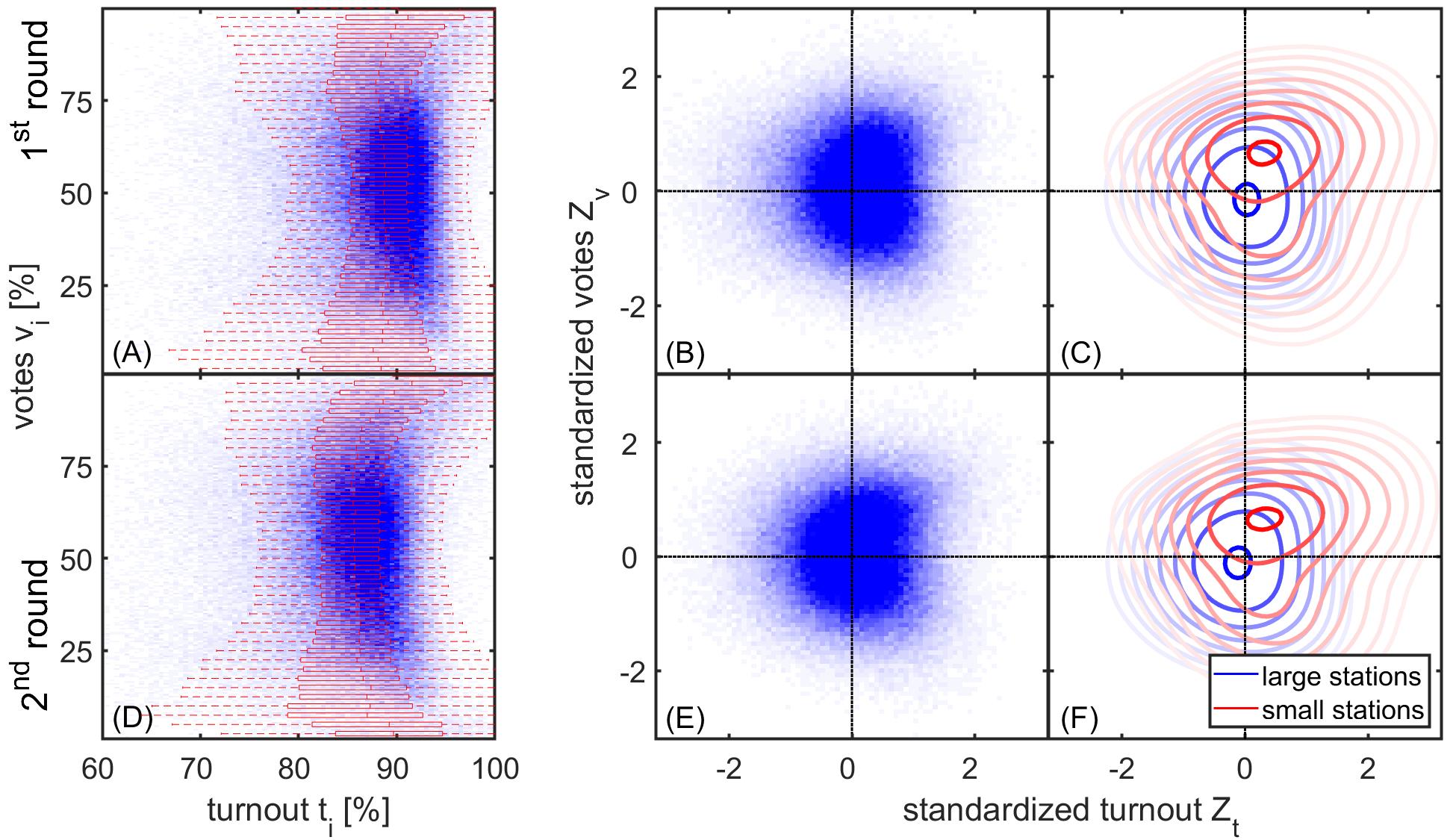}
\end{center}
 \caption{Forensic electoral fingerprints for the two rounds of the 2023 presidential election. The fingerprints for (A) first round and (B) second round show joint vote-turnout distributions with color intensity encoding the number of ballot boxes with a given vote (y-axis) and turnout (x-axis). For both elections, we find a visible correlation in the region of high vote and turnout (e.g. more than 80\%), which can be associated with ballot stuffing. A box plot (red horizontal boxes) shows the 25th, 50th and 75th percentiles of voter turnout as a function of votes (whiskers indicate the 95\% confidence interval). (C) To adjust for regional characteristics, the fingerprints can be adjusted by rescaling the vote and turnout shares by their typical levels in the unit's region, resulting in the standardised fingerprint shown. (D) Traces of voter coercion can be identified by comparing the standardised fingerprints of small (red lines) and large (blue) units, as voter coercion results in their being shifted towards inflated votes and turnout, as observed here. (E,F) The  standardised fingerprints of the second round are similar to those of the first round.}
 \label{FP}
\end{figure*}

In Figure~\ref{FP}, we test for the presence of electoral malpractices that lead to vote-turnout correlations (such as ballot stuffing) using so-called electoral fingerprints, i.e., a 2-d histogram of the vote-turnout distribution. The fingerprint for the first round and second round of the 2023 Turkish presidential election is shown in Figure~\ref{FP}(A,D), respectively. The colour intensity (blue) indicates the number of boxes with the corresponding percentage of votes (x-axis) and turnout (y-axis), together with a box plot of the distribution of turnout for a given percentage of votes. If there were no non-linear correlations between votes and turnout, the bulk of the distribution in Figure~\ref{FP}(A,D) should be circular or elliptical symmetric. Malpractices such as ballot stuffing would inflate turnout and simultaneously increase vote shares, breaking the elliptical symmetry in the fingerprints if the number of affected boxes is large enough.

Considering the region of high voting and turnout, in both rounds we see a smearing of the bulk towards inflated votes and turnout, towards values of 100\% votes for Erdo\u{g}an and 100\% turnout. To assess whether such deviations between symmetric and biased fingerprints are statistically significant, we run a parametric test that was proposed previously \cite{Klimek18}. This model is designed to test if the observed deviations from the normal distribution in vote and turnout shares can be better explained by a model where ballot stuffing occurs in a given fraction of ballot boxes (fraud parameter $f$). To fit and evaluate this statistical model, we follow a previously described strategy \cite{Klimek18} and further restrict the analysis to boxes with a vote and turnout share of more than 25\%.

For the first round we find a fraud parameter of $f=0.024$ with a standard deviation (SD) of $0.019$ and for the second round $f=0.19$ (SD $1.7\%$).
Note that for 2018 we find higher values with $f=0.085$ (SD $0.039$). Therefore, the test suggests that the number of ballot boxes affected by such statistical irregularities has decreased from 2018 to 2023, to a point where the ballot stuffing test does not detect statistically significant effects.

\subsection{Voter rigging test.}
The fingerprints shown in Figure~\ref{FP}(A) and (D) may also show deviations from elliptical symmetry due to geographical effects. To account for such effects, it has been suggested to compare the unit with other units in close geographical proximity \cite{Jimenez17}. Here, we compare the vote and turnout figures of a polling station with the averages observed in other polling stations in the same constituency. We refer to these rescaled vote and turnout shares as standardised votes and turnout, respectively. We call their joint distribution (2D histogram) the ``standardised fingerprint''
.
Standardised fingerprints are shown in Figure~\ref{FP}(C) for round 1. For the voter coercion test, we ask whether small and large units have different standardised fingerprints. The underlying hypothesis of this test is that coercion is more likely to occur in smaller units because they are more susceptible to coercion tactics. Reasons for this include that (i) it is easier to identify opponents in smaller units, (ii) fewer eyewitnesses can be expected, and (iii) election observers are less likely to be present. In line with these assumptions, voter manipulation suggests that the standardised fingerprints of small units are biased towards increased voting and turnout compared to larger units.

We use different definitions of ``small units". Figure~\ref{FP}(B) shows the standardised fingerprints for small (red) and large (blue) units, where small units are those in the lowest $p=10$th percentile of all units. It can clearly be seen that the fingerprints for small units are shifted towards the upper right corner, see Figure~\ref{FP}(C), which is consistent with voter manipulation. For the second round, we found almost identical standardised fingerprints, see Figs. 2(E) and 2(F).

\begin{figure}[tbp]
\begin{center}
 \includegraphics[width = 0.45\textwidth, keepaspectratio = true]{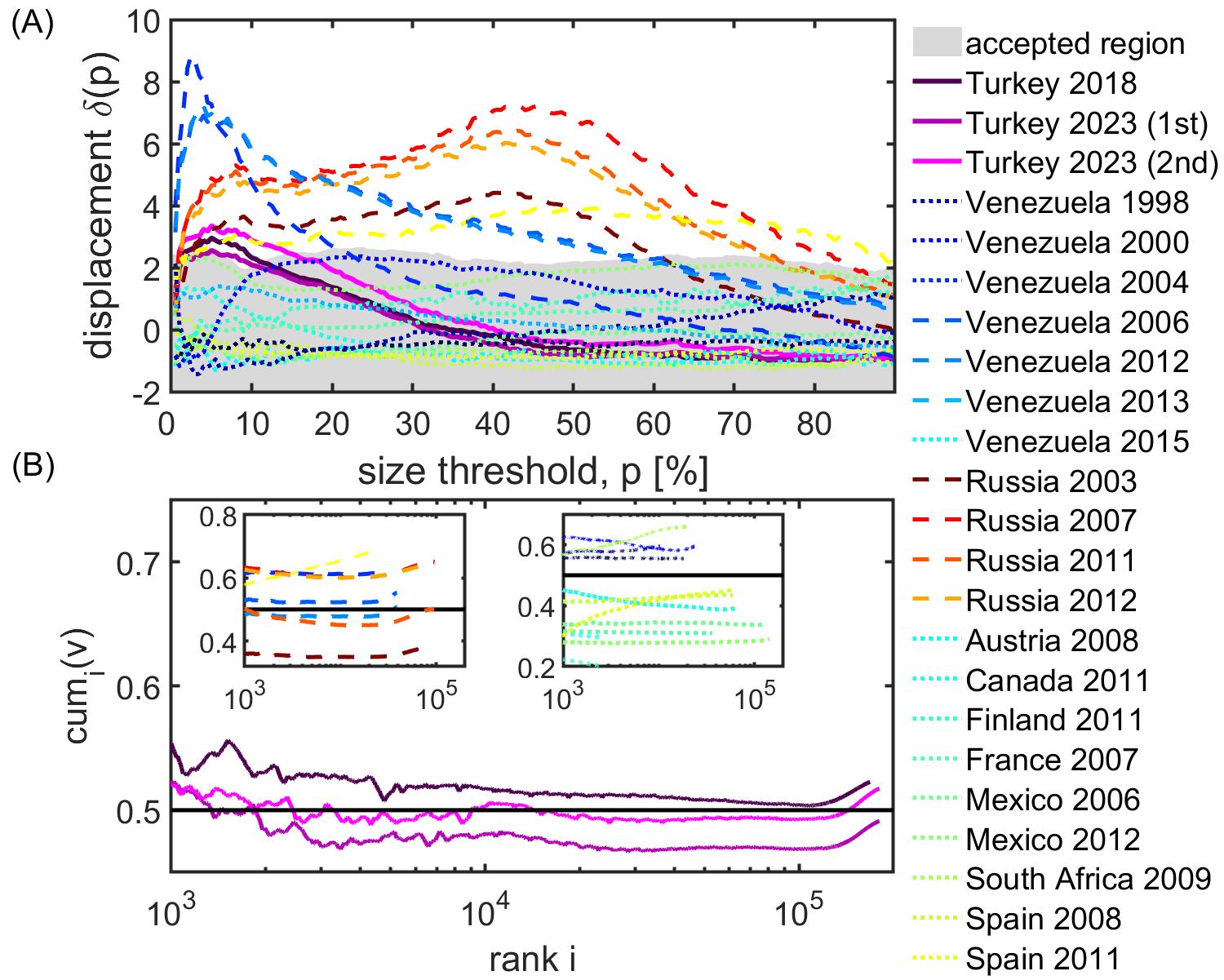}
\end{center}
 \caption{Results of the statistical test for voter manipulation. (A) The displacement $\delta(p)$ between small and large units for the first (solid dark magenta line) and second (solid light magenta line) round of the 2023 elections is very narrowly outside the accepted range for a restricted set of size thresholds, similar as it was in 2018 (solid black line). These displacements are much smaller displacements in the Russian or Venezuelan elections (dashed lines); the reference elections are shown as dotted lines. (B) Units in the 2023 and 2018 Turkish elections are ranked according to their electorate size. We show the cumulative vote share, $cum_i(v)$, calculated over all units with a size greater than the given rank. As in 2018, we observe a characteristic ``hockey stick'' in 2023, meaning that units with high rank (low electorate size) show a clear tendency to favour Erdo\u{g}an.}
 \label{ES}
\end{figure}

The magnitude of the displacement between the average standardised votes and the turnout of small and large units depends on the size threshold p and is denoted by $\delta(p)$, see \cite{Klimek18, Jimenez17} for methodological details. To assess whether this shift is statistically significant, we apply the Jimenez et al. voter rigging test \cite{Jimenez17}. The idea behind this test is to estimate the expected shifts between small and large units based on a reference set of trustworthy elections, yielding a range of ``acceptable shift sizes". We obtain this acceptable region from 21 different reference elections in ten countries, see \cite{Klimek18, Jimenez17}. For a given election, one can now check whether the actual observed displacement between small and large units for a size threshold $p$ lies within this region (``accepted region'') or not.

The displacement, $\delta(p)$ , is shown in Figure~\ref{ES}(A) for the reference set of elections (solid lines), the elections in Russia and Venezuela (dashed line) next to the two rounds in 2023 (solid magenta) and 2018 (solid black) Turkish elections. For small size thresholds,$p$, all Turkish datasets show shifts that are slightly outside the acceptable range. This indicates statistically significant signs of voter manipulation, however, only for a limited region of thresholds. The shifts in the first round in 2023 are slightly smaller than in 2018, whereas in the second round they were slighly larger. 

To assess the potential impact of these voter rigging effects in the data, we rank the units by their electorate size in descending order and calculate the vote share over all units with smaller ranks (higher electorate size), see Figure~\ref{ES}(B). In this plot, voter manipulation takes the form of a ``hockey stick'', i.e., a sharp increase for the smallest units. This signal is also found in Russia and Venezuela, but is absolutely absent in the reference elections, see the insets in Figure~\ref{ES}(B).

\begin{figure*}[tbp]
\begin{center}
 \includegraphics[width = 0.9\textwidth, keepaspectratio = true]{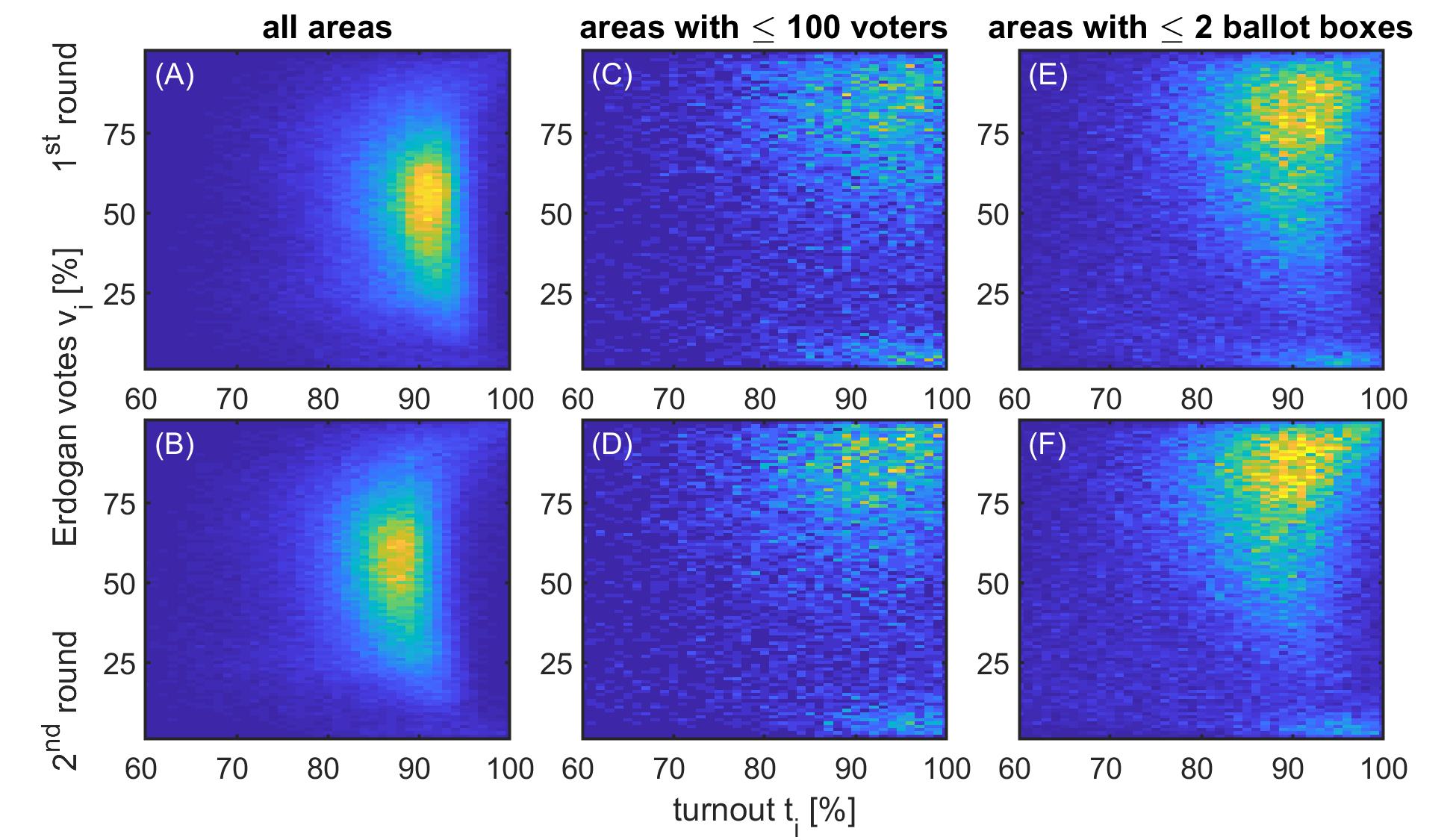}
\end{center}
 \caption{The fingerprints in 2023 for small units (C-F) show very different patterns than those with an electorate size of more than 100 (A,B). Considering only areas with an electorate size of less than 100 (C,D), or alternatively ballot boxes from areas with one or two ballot boxes (E,F), results in a bimodal distribution with a large mode in the region with very high turnout and votes for Erdo\u{g}an.}
 \label{SFP}
\end{figure*}

To further show the bias in small units in the first and second round in 2023, respectively, we compare the fingerprint observed in all units with an electorate size larger than 100, Figure~\ref{SFP}(A,B), with the fingerprints of two different definitions of ``small". First, we consider only boxes with an electorate of one hundred or less, resulting in 11,689 units, see Figure~\ref{SFP}(C,D). Alternatively, one can consider only boxes from areas with one or two ballot boxes in total, resulting in 38,662 boxes in the first round, see Figure~\ref{SFP}(E,F). The small units show completely different fingerprints when compared to the large ones. They show a bimodal distribution with a larger mode in the high vote-high turnout region and a smaller mode in the low vote-high turnout region.

\begin{figure}[tbp]
\begin{center}
 \includegraphics[width = 0.45\textwidth, keepaspectratio = true]{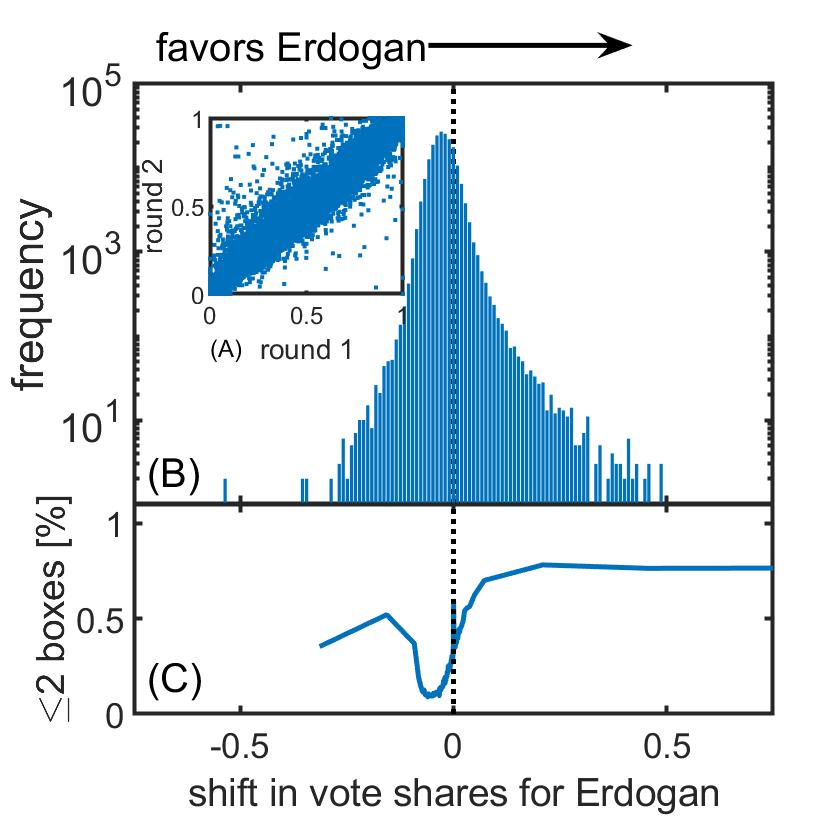}
\end{center}
 \caption{Statistical analysis of vote shifts from Kili\c{c}daro\u{g}lu to Erdo\u{g}an from the first to the second round. (A) The vote shares for Erdo\u{g}an and O\u{g}an in the first round are strongly correlated with Erdo\u{g}an's vote share in the second round, although there are some outliers. (B) A histogram of these vote shifts shows that these outliers typically favour Erdo\u{g}an in the sense that large shifts from anti to pro Erdo\u{g}an are much more common than the other way around. (C) Looking at the proportion of ballot boxes coming from districts with two or fewer ballot boxes as a function of vote shifts, we see that these extreme vote shifts favouring Erdo\u{g}an occur predominantly in districts with fewer ballot boxes.}
 \label{VS}
\end{figure}

Comparing Figures~\ref{SFP}(E) and (F), there is a tendency for the high vote/high turnout mode to have shifted further to the upper right of the plot in the second round compared to the first.
To systematically compare such effects, we examine the shift of vote shares between Erdo\u{g}an or O\u{g}an in the first round and Erdo\u{g}an in the second round (see Methods).
A scatterplot of these vote shares is shown in Figure~\ref{VS}(A), which shows that in the vast majority of cases the vote shares for Erdo\u{g}an (or candidates who later endorsed him) were similar in the first and second rounds; the points cluster around the $x=y$ line.
The distribution of vote shifts, $\delta v_i$, in Figure~\ref{VS}(B) shows a clear tendency for large shifts in vote shares from Kili\c{c}daro\u{g}lu in round 1 toward Erdo\u{g}an in round 2 to be more common than vice versa (the distribution is skewed to the right, favouring Erdo\u{g}an).
Figure~\ref{VS}(C) shows the percentage of ballot boxes that are in regions with two or fewer ballot boxes in total for each of the bins in the histogram (bins in the lowest or highest percentile have been combined).
Interestingly, larger districts (in the sense of having more than two ballot boxes) have very small vote shifts.
Strong positive vote shifts in favour of Erdo\u{g}an occur mainly in districts with few ballot boxes.
  
Similar observations apply, to a lesser extent, to vote shifts in favour of Kili\c{c}daro\u{g}lu.
To compare the magnitude of these vote shifts, we consider a symmetrized distribution of vote shifts in which vote shifts in favour of Erdo\u{g}an follow the same distribution as those in favour of Kili\c{c}daro\u{g}lu (see Methods).
Overall, we find that Erdo\u{g}an received 342,000 (SD 4,900) excess votes, which corresponds to 0.64\% of all valid votes.

\section{Discussion}

An electoral forensic analysis of the first and second round of the 2023 presidential election in Turkey identifies statistical irregularities similar to those observed in the 2018 election and the 2017 constitutional referendum. However, the estimated magnitude of these irregularities has decreased in 2023 compared to the 2018 presidential election. For 2023, we observe trends in turnout inflation, as would be expected in the presence of electoral malpractices such as ballot stuffing. However, the percentage of electoral units potentially affected by these distortions has fallen to 2.4\% (SD 1.9\%) in the first round and to 1.9\% (SD 1.7\%), making the results statistically insignificant.
In both rounds in 2023, we also observe a tendency for areas with small electorates to show different voting and turnout patterns compared to other regions. Such biases are consistent with the presence of voter coercion or intimidation techniques, to which smaller and more remote electoral units are more susceptible. The effect size of these deviations is statistically significant only by a small margin and for a limited range in the parameter space (meaning these effects vanish when increasing the size threshold of what constitutes ``small'' unit above the 20th percentile).

We found, however, suprising irregularities when considering ballot boxes in areas with two or fewer ballot boxes in total, which typically hints at remote areas.
There are roughly 40,000 such boxes (out of 190,000 in total) that show completely different trends in their vote--turnout distribution when compared to the other areas.
In such remote areas the vote--turnout distribution becomes bimodal with high vote/high turnout bulks for both candidates (Erodgan and Kili\c{c}daro\u{g}lu, respectively), though the Erdo\u{g}an mode contains much more ballot boxes compared to the Kili\c{c}daro\u{g}lu mode.
What is more, comparing the first and the second round of the election, a much higher fraction of such boxes has flipped from the Kili\c{c}daro\u{g}lu to the Erdo\u{g}an mode compared to the other way around.
This high number of districts (around 90\%) that flipped from one candidate to the other at very high turnout levels in the span of no more than two weeks is certainly surprising.
Correcting the election results for such suprising vote shifts would reduce Erdo\u{g}an's vote tally by about 342,000 votes or 0.64\% of valid votes.

When interpreting the results of forensic voting tests, several limitations must be borne in mind.
First and foremost, none of these tests can provide incontrovertible evidence of electoral fraud; they provide correlation, not causation.
Although most of the tests make adjustments for regional characteristics and rule out the possibility that these correlations are spurious artefacts from small districts, it is not possible to control for all potential confounders.
A positive result in such a test indicates that the data are compatible with certain types of fraud (ballot stuffing, etc.) and typically gives a quantitative estimate of how many regions might have been affected by such malpractice.
Other non-fraudulent influences can never be completely ruled out.
In order to understand whether the observed irregularities may be due to such non-fraudulent phenomena, such as heterogeneous voter mobilisation through strategic voting, forensic tests must always be evaluated in conjunction with external information \cite{Mebane16,Mebane18}.
Conversely, even if the tests give a negative result, one cannot rule out the presence of other types of fraud, which would require a different test.

Taken together, these results suggest that the presence of certain types of electoral malpractice in both rounds of the 2023 presidential election cannot be ruled out. However, these malpractices appear to have been less frequent in 2023 than in 2018, and less decisive in swinging the vote one way or the other. 
There is a notion that Turkish elections have become ``free and unfair'' because Turkey's political playing field is known to be massively tilted in Erdo\u{g}an's favour \cite{WaPo23a}.
The authors are satisfied that Erdo\u{g}an or his supporters have autocratised the media, the judiciary, civil society and academia \cite{Coskun20, Esen20}.
Our analysis suggests that it may be more appropriate to describe Turkish elections as ``mostly free and unfair'', as consistent trends of small but discernible electoral irregularities can be consistently found by electoral forensic tools.
While these statistical irregularities were not large enough to determine the outcome in 2023 on their own, they certainly tilt Turkey's political playing field even further towards an illiberal democracy.

\end{document}